# Heat capacity and MÖssbauer study of Self flux grown FeTe Single Crystal

P. K Maheshwari[1,2], V. Raghavendra Reddy[3] and V.P.S. Awana[1*]

[1]*CSIR-National Physical Laboratory, Dr. K. S. Krishnan Marg, New Delhi-110012, India*
[2]*AcSIR-National Physical Laboratory, New Delhi-110012, India*
[3]*UGC-DAE Consortium for Scientific Research, Khandwa Road, Indore-425001*

ABSTRACT

We report mainly the heat capacity and Mössbauer study of self flux grown FeTe single crystal, which is ground state compound of the Fe chalcogenides superconducting series, i.e., $FeTe_{1-x}(Se/S)_x$. The as grown FeTe single crystal is large enough to the tune of few cm and the same crystallizes in tetragonal structure having space group of P4/nmm. FeTe shows the structural/magnetic phase transition at 70K in both magnetic and resistivity measurements. Heat capacity measurement also confirms the coupled structural/magnetic transition at the same temperature. The Debye model fitting of low temperature (below 70K) heat capacity exhibited Debye temperature ($\Theta_D$) to be 324K. Mössbauer spectra are performed at 300K and 5K. The 300K spectra showed two paramagnetic doublets and the 5K spectra exhibited hyperfine magnetic sextet with an average hyperfine field of 10.6Tesla matching with the results of Yoshikazu Mizuguchi et al.

***Corresponding Author**
Dr. V. P. S. Awana,
Principal Scientist
E-mail: awana@nplindia.org
Ph. +91-11-45609357, Fax-+91-11-45609310
Homepage awanavps.wenbs.com

INTRODUCTION

Fe based superconductors, firstly in Fe pnictides [1-2] followed by Fe chalcogenides [3-4], engrossed huge attention of the scientific community in condensed matter physics. These Fe based superconductors fall in the high temperature superconductivity category, as their superconducting transition temperature ($T_c$) is reported to be as high as 56K at ambient pressure [2,5]. Like HTS Cuprates [6,7], the Iron based superconductors also do not follow the conventional electron phonon theory of superconductivity. The understanding of the ground state of the non conventional superconductors (Cuprates and Fe pnictides/chalcogenides) is as important as appearance of superconductivity in them after doping of carriers. Interestingly, the ground state of both mysterious superconducting Cuprates and Fe pnictides/chalcogenides is magnetic in nature [1-9]. There is a broad consensus that the magnetic order disappears before onset of superconductivity in these systems.

FeTe, is not a superconductor, but have anti-ferromagnetic (AFM) ordering below70K[8], which becomes superconducting by Te site doping of Se/S with $T_c$ of up to 15K [4,9-11].The superconductivity in Fe chalcogenides have been increased up to 50K with intercalation of favorable alkali metals [12]. While the FeTe, parent compound of Iron chalcogenides shows the drastic changes in electrical resistivity as well as in magnetic susceptibility at 70K, which belong to anti-ferromagnetic ordering along with first order structural phase transition [8]. This first order phase transition is seen clearly in heat capacity measurement at 70K. $T_c$ of optimally doped superconducting iron chalcogenides i.e., $FeTe_{1/2}Se_{1/2}$ rises up to 37K with applying external pressure [13, 14], although no superconducting phase occurs for un-doped $FeTe_{0.92}$ under high pressure of even up to 19GPa [15].

Concerning the fact that the ground state of mysterious Cuprates and Fe pnictides/chalcogenides is as elusive as their superconductivity, we choose to focus on Fe chalcogenides ground state i.e., FeTe. One of the basic probes to characterize the magnetic states, particularly in Fe based compounds is Mössbauer spectroscopy, which is very sensitive tool concerning the hyperfine magnetic field etc. and different Fe coordination states respectively below and above the magnetic ordering temperature [16].

Although good around a decade is passed after discovery of Fe pnictides/chalcogenides superconductors, the improvements in quality of the material is yet in progress. In particular the quality large enough single crystals are in very much demand. Generally, travelling floating zone/ Bridgman technique has been used for growth of iron chalcogenides single crystal [17,18]. In some cases, the flux free method is also used to synthesize Iron chalcogenides single crystals, but the crystals grown were tiny [19-21]. Recently, we succeeded in growth of flux free large single crystals of FeTe and presented its synthesis and structural properties [22]. In, present letter, we report the heat capacity and Mössbauer study analysis of self flux grown FeTe.

EXPERIMENTAL DETAILS

FeTe single crystal was synthesized by flux free method via solid state reaction route in a simple automated furnace. Details of crystal growth of studied FeTe single crystal are reported earlier [22]. X-ray diffraction pattern has been performed at room temperature using Rigaku X-Ray diffractometer, having wave length of 1.54184Å of $CuK_\alpha$ line. Scanning electron microscopy (SEM) images and EDAX analysis were performed on a ZEISS-EVO MA-10 scanning electron microscope. Transport and thermal measurement has been performed from room temperature to 5K using Quantum Design Physical Property Measurement System (PPMS)-140 kOe down to 2K. Mossbauer measurements are carried out on the crushed powder sample of the prepared single crystals. $^{57}$Fe Mossbauer measurements were carried out in transmission mode with $^{57}$Co radioactive source in constant acceleration mode using standard PC-based Mossbauer spectrometer equipped with WissEl velocity drive. Velocity calibration of the spectrometer is done with natural iron absorber at room temperature. Low temperature $^{57}$Fe Mossbauer measurements are carried out using Janis make superconducting magnet. The spectra are analyzed with NORMOS program.

RESULTS AND DISCUSSION:

Figure 1 shows the single crystal X-Ray Diffraction pattern of synthesized FeTe single crystal at room temperature. From this XRD pattern, it may be concluded that the growth of the synthesized single crystal is in (00*l*) plane, as high peak intensity occurs in (001), (002), (003) and (004) planes only. *Inset* view of Fig 1 shows the fitted powder XRD pattern after crushing the synthesized sample at room temperature using Rietveld refinement. The Rietveld refinement analysis concluded that studied FeTe sample crystallizes in tetragonal structure with P4/nmm space group and lattice parameters *a*=*b*=3.82(2)Å, *c*=6.29(1)Å. Detailed analysis of XRD of FeTe single crystal had been reported earlier [22].To understand the morphology and quantitative analysis of FeTe single crystal, SEM and EDAX had been performed at room temperature. SEM image of FeTe has been shown in figure 2 and *inset view* of figure 2 is EDAX result of the same. SEM result shows the layered structure, indicating single crystalline morphology of the sample. EDAX results show the elemental analysis of FeTe compound, confirming near stoichiometric ratio of 1:1 ratio. Further, more detailed analysis of SEM and EDAX had been reported earlier [22].

Figure 3 shows the electrical resistivity (ρ) verses temperature (*T*) measurement of FeTe from 300K to 5K in cooling and warming cycle protocols, without applying external magnetic field. From the ρ(*T*) measurement, we can conclude that FeTe shows the semiconducting behavior during cooling cycle till structural/magnetic phase transition occurring at around 66K and the same is step like metallic below this temperature. The same structural/magnetic phase transition occurred at 66K during cooling cycle and 71K during warming cycle for FeTe single crystal. So we can conclude that there is hysteresis ($\Delta T$) of 5K during cooling and warming cycle, this hysteresis occurs due to presence of latent heat in the sample. This structural/magnetic phase transition and hysteresis ($\Delta T$) also occurs in magnetization measurement, which is shown in *inset view* of figure 3. The DC magnetic susceptibility verses temperature plot from temperature range from 150K to 5K during cooling and warming cycle for FeTe single crystal exhibited hysteresis and the plot is AFM (anti-ferromagnetic) like. The DC magnetic susceptibility measurement showed the sharp transition being occurring approximately at same temperature as in resistivity measurement. More detailed analysis of transport and magnetic measurements had been reported earlier [22].

Figure 4 shows the between $C_p$ verses temperature plot for FeTe single crystal. In the heat capacity measurement, there is sharp spike like structural transition occurring nearly at 70K, which confirms the structural/magnetic phase transition in FeTe sample. This transition occurs due to heat absorption and release in the sample, leading to first order phase transition. This first order phase transition provides sharp changes in entropy during cooling in FeTe sample, suggesting the fact that change in entropy also plays an important role during phase transition [22]. $C_p(T)$ verses $T$ curve has been fitted using Debye model at low temperature (below magnetic/structural transition) from 70K to 5K. The fitted curve using Debye equation $C_p(T)= \gamma T + \beta T^3 + \delta T^5$ has been shown in inset view of figure 4, where γ, β and δ are Debye constants. The calculated value of Debye constants γ, β and δ are 31.36mJ/mol-K$^2$, 0.114mJ/mol-K$^4$ and -4nJ/mol-K$^6$ respectively after fitting the curve using Debye equation, which are near to earlier reported result [23]. Debye temperature ($\Theta_D$) is calculated using Debye model relation i.e. $\Theta_D=(12\pi^4 N_A z k_B/5\beta)^{1/3}$, where $N_A$ (Avogadro constant) = 6.023 x $10^{23}$ mol$^{-1}$, z= Numbers of atom per unit i.e. 2 for FeTe and $k_B$ is Boltzmann constant (=1.38 x $10^{-23}$m$^2$kg s$^{-2}$K$^{-1}$). Debye temperature is found to be 324K for FeTe single crystal using Debye model.

Figure-5(a) and 5(b) show the Mossbauer spectra of the as prepared sample at 300K and 5K respectively. The room temperature data shows an asymmetric paramagnetic doublet indicating the presence of two Fe sites similar to that of Yoshikazu Mizuguchi et al [24]. The room temperature data is fitted with two paramagnetic doublets and the isomer shift and quadrupole splitting values are 0.43 ± 0.01, 0.16 ± 0.01mm/s for the first doublet with a spectral area of 38% and 0.46 ± 0.01, 0.43 ± 0.01mm/s for the second doublet with a spectral area of 62%. The main doublet corresponds to paramagnetic Fe of FeTe layers, while the second one belongs to excess Fe ions at interlayer site [24]. At 5K, the Mossbauer spectra shows magnetic sextet. This is consistent with the fact that the prepared sample exhibits magnetic ordering below 70 K [24]. The data is fitted with distribution of hyperfine fields and the obtained distribution is shown as inset of Figure-5(b). An average hyperfine field of about 10.6 Tesla is obtained matching with that of Yoshikazu Mizuguchi et al [24].

In conclusion, we synthesized FeTe single crystal using flux free method via solid state reaction. The FeTe single crystal is having crystallite of tetragonal structure with P4/nmm space group. In FeTe, structural/magnetic phase transition occurs nearly 66K and 72K during cooling and warming cycle in both resistivity and magnetic measurements. In heat capacity measurement, we observed a sharp spike is occurred at nearly 70K, which is the confirmation of structural/magnetic phase transition occurring during the cooling cycle. Heat capacity curve is fitted in low temperature using Debye model from 65K to 5K, and Debye temperature is found nearly 324K. Mössbauer spectra showed paramagnetic doublets at room temperature and hyperfine magnetic sextet at 5K.

ACKNOWLEDGEMENT:

Authors would like to thank Director NPL-CSIR India for his intense interest in the present work. This research work is financially supported by *DAE-SRC* outstanding investigator award scheme on search for new superconductors. P. K. Maheshwari also thanks to CSIR, India for research fellowship and AcSIR-NPL for Ph.D. registration.

**Figure Captions**

**Figure 1:** FeTe single crystal XRD at room temperature. Inset is fitted powder XRD of same using Rietveld refinement at room temperature.

**Figure 2:** SEM image of FeTe single crystal at room temperature. Inset is EDAX quantitative analysis of same.

**Figure 3:** Electrical resistivity measurement from temperature range of 300K to 5K during cooling and warming cycle for FeTe single crystal. Inset is DC magnetic susceptibility measurement at 1000 Oe from temperature range of 150K to 5K for FeTe single crystal.

**Figure 4:** The temperature dependent heat capacity $C_p(T)$ in temperature range 250-5K of FeTe single crystal.

**Figure 5:** Mössbauer spectra of powdered FeTe single Crystal at the indicated temperatures viz., 5(a) 300K and 5(b) 5K. The inset of fig 5(b) shows the distribution of hyperfine fields.

Figure 1:

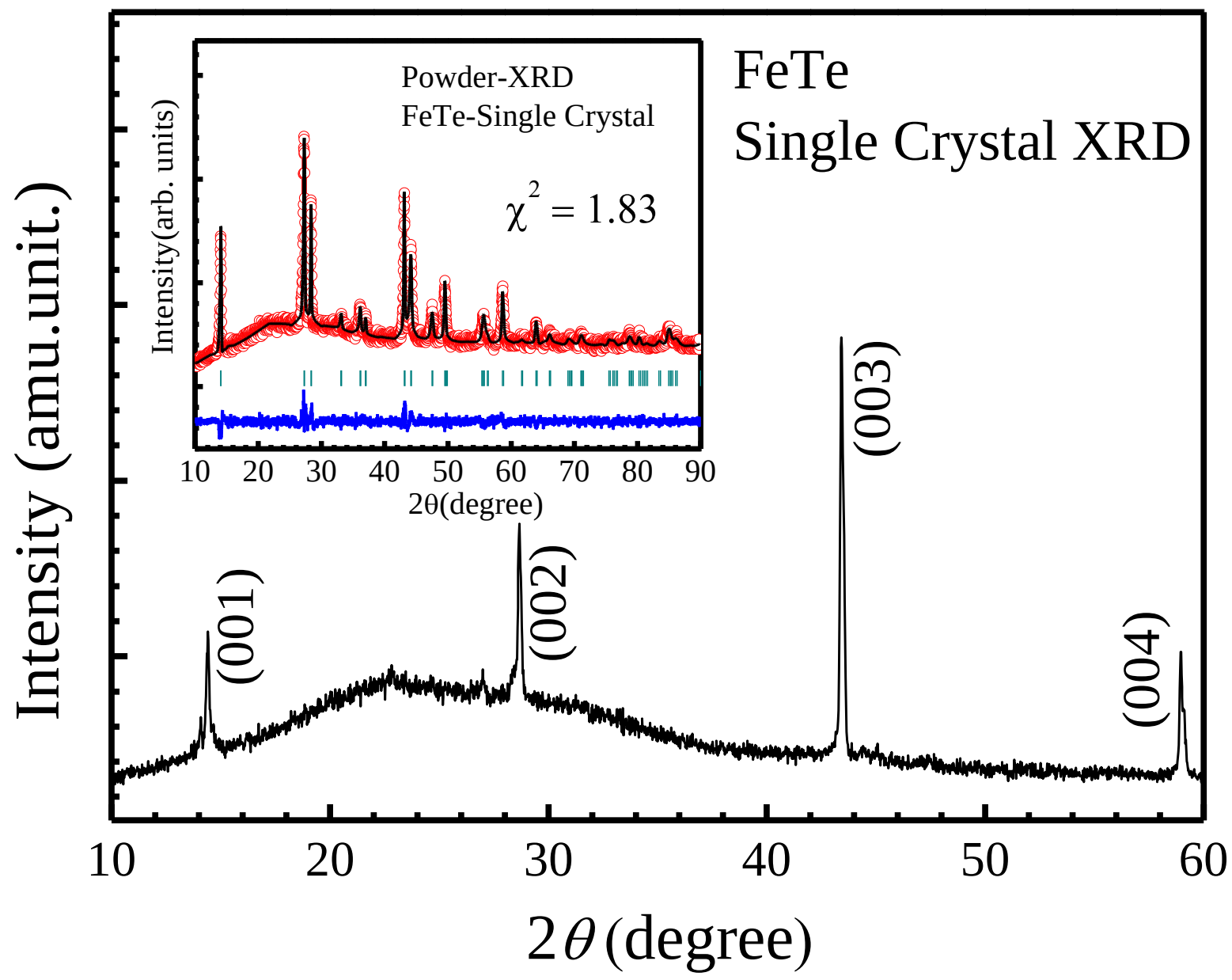

Figure 2:

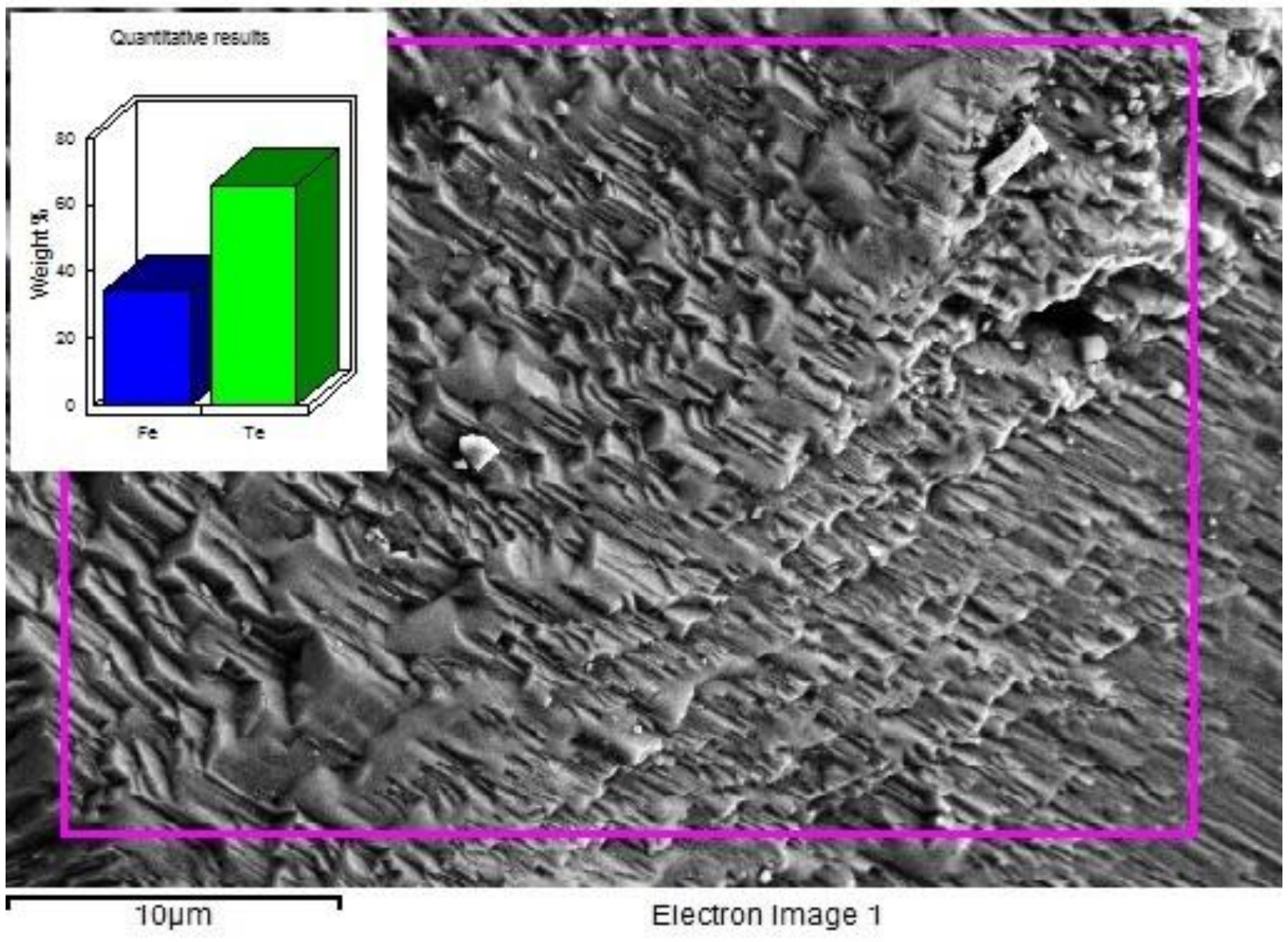

Figure 3:

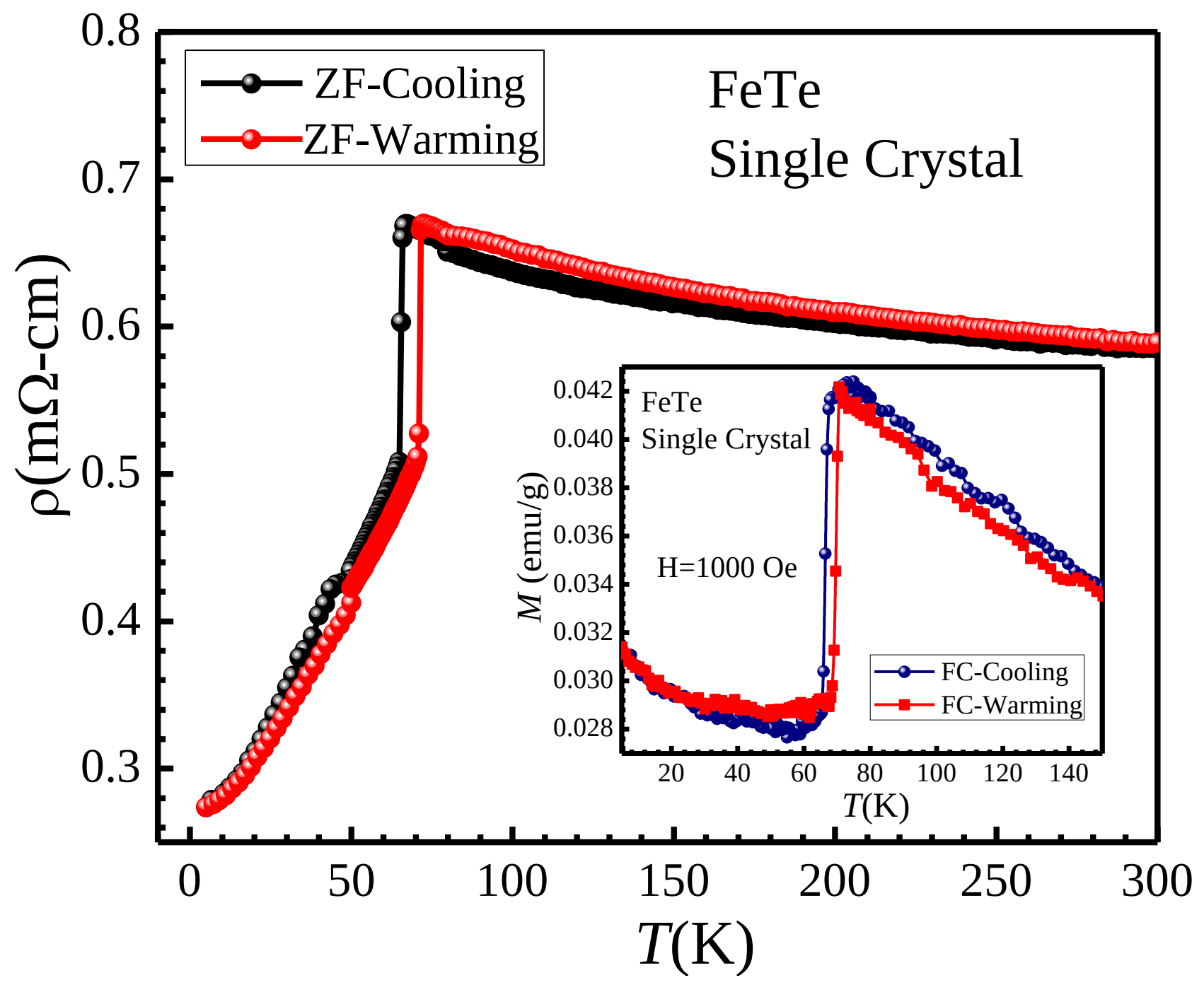

Figure 4:

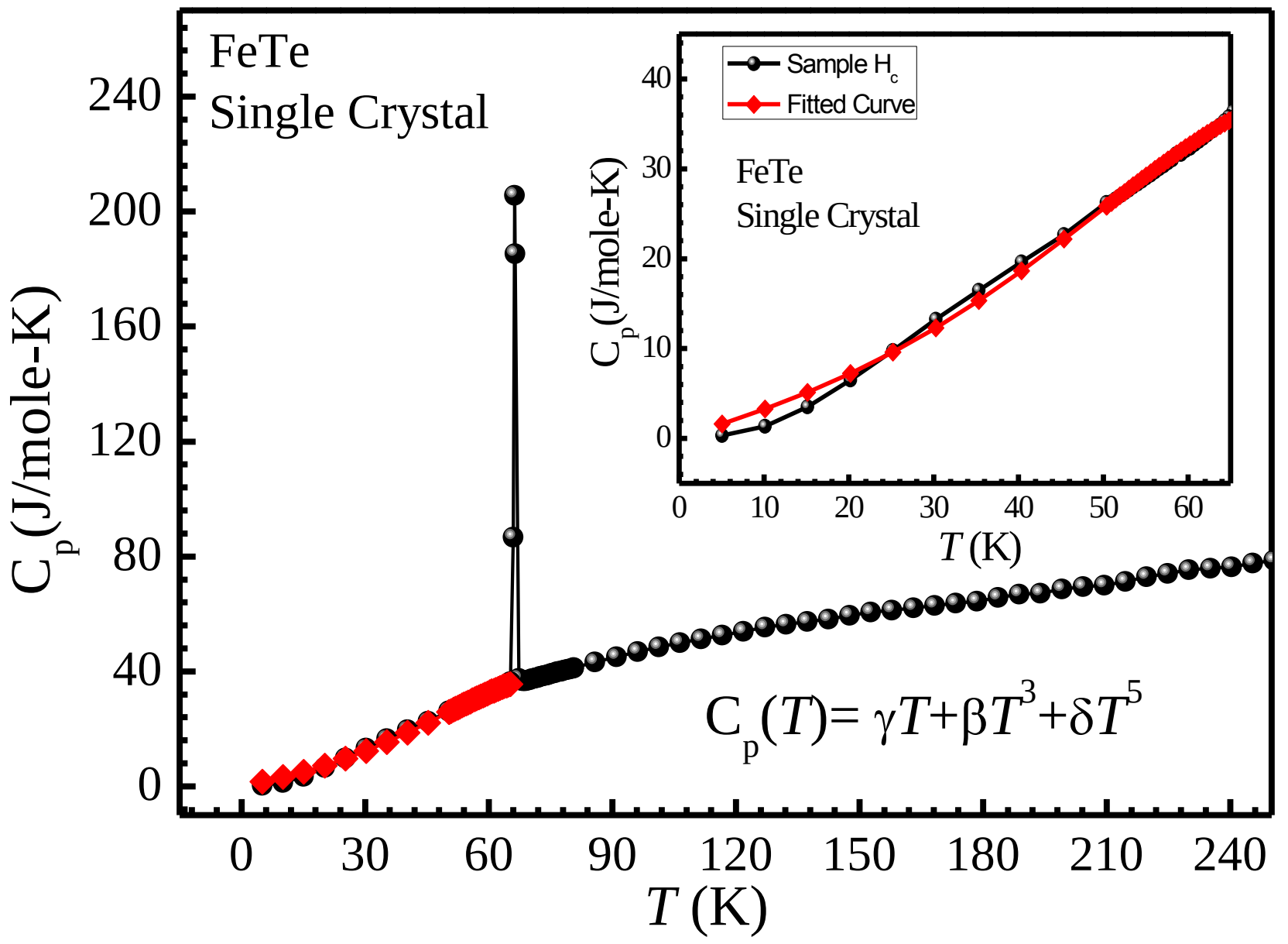

Figure 5(a)

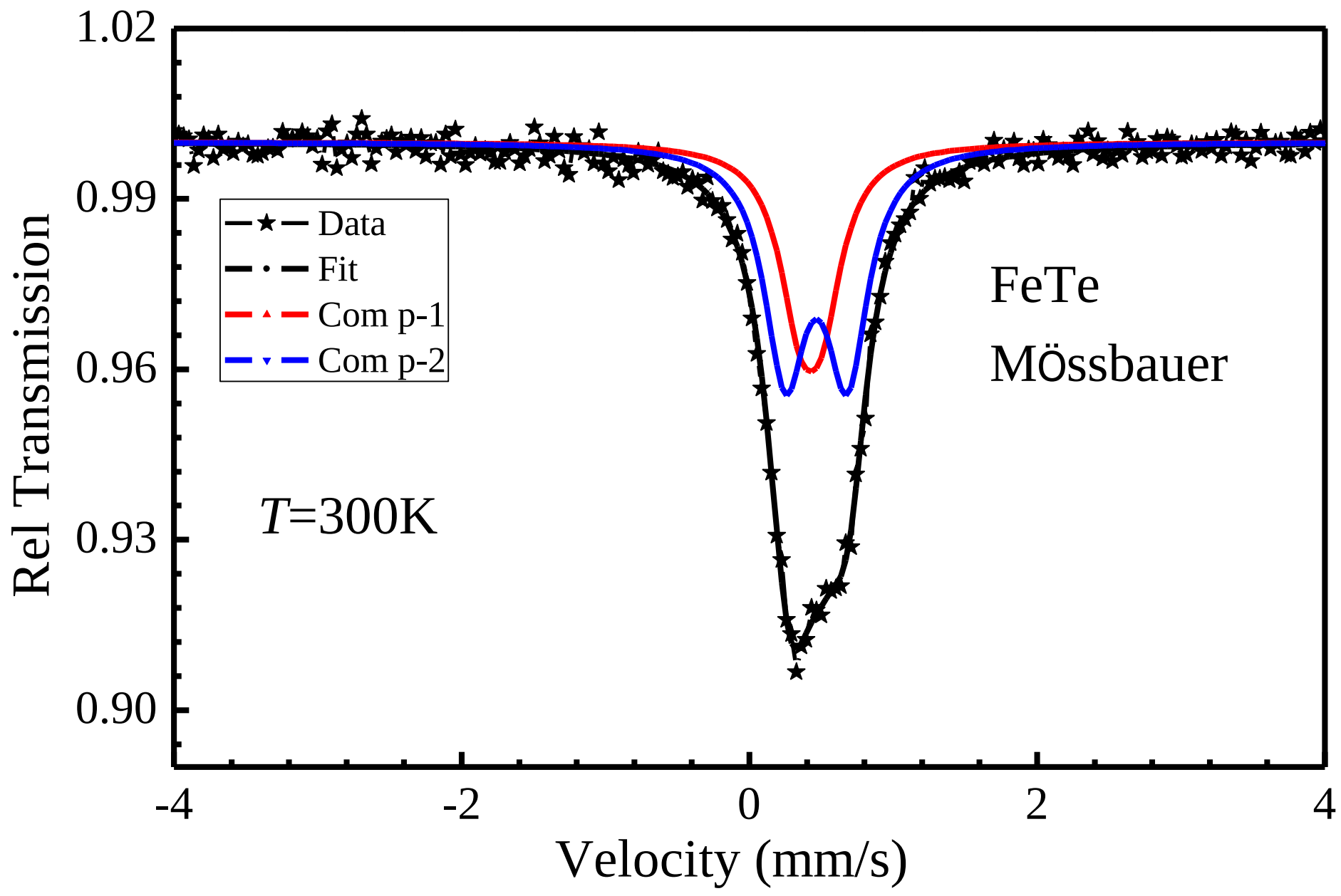

Figure 5(b)

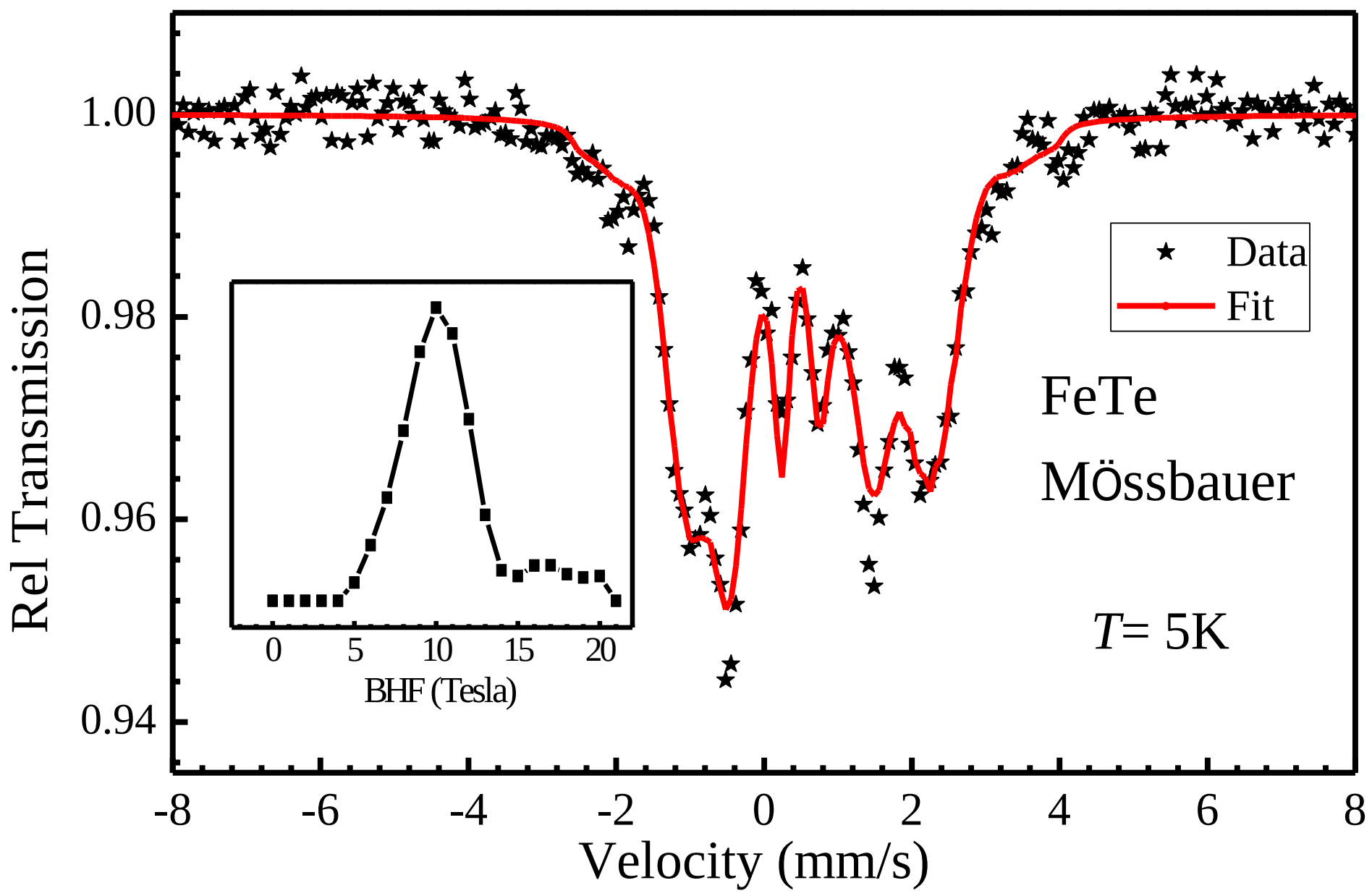